\documentclass{elsart}
\usepackage{amsfonts}
\usepackage{graphicx}

\newcommand{\wb}{\omega_{\mathrm{b}}}
\newcommand{\wt}{\omega_{\mathrm{trap}}}
\newcommand{\we}{\omega_{\mathrm{e}}}
\newcommand{\wph}{\omega_{\mathrm{ph}}}
\newcommand{\wo}{\omega_0}
\newcommand{\w}{\omega}
\newcommand{\Ui}{U_\mathrm{in}}
\newcommand{\Uo}{U_\mathrm{out}}
\newcommand{\Ki}{K_\mathrm{in}}
\newcommand{\Ko}{K_\mathrm{out}}
\newcommand{\Ut}{U_\mathrm{trap}}
\newcommand{\Uph}{U_\mathrm{ph}}
\newcommand{\ii}{\mathrm{i}}
\newcommand{\F}{\mathcal{F}}

\begin{document}

\begin{frontmatter}

\title{Discrete moving breather collisions in a Klein-Gordon chain of oscillators}

\author{A Alvarez and FR Romero\thanksref{FRR}}
\address {Grupo de F\'{\i}sica No Lineal. \'{A}rea de F\'{\i}sica
Te\'orica. Facultad de F\'{\i}sica. Universidad de Sevilla. Avda.
Reina Mercedes, s/n. 41012-Sevilla (Spain)}
\author{J Cuevas and JFR Archilla}
\address{Grupo de F\'{\i}sica No Lineal. Departamento de Fisica
Aplicada I. ETSI Inform\'{a}tica. Universidad de Sevilla. Avda.
Reina Mercedes, s/n. 41012-Sevilla (Spain)}

\date{27 June 2007}
\journal{Physics Letters A}
\thanks[FRR]{Corresponding author. E-mail: romero@us.es}


\begin{abstract}

We study collision processes of moving breathers with the same
frequency, traveling with opposite directions within a Klein-Gordon
chain of oscillators. Two types of collisions have been analyzed:
symmetric and non-symmetric, head-on collisions. For low enough
frequency the outcome is strongly dependent of the dynamical states
of the two colliding breathers just before the collision. For
symmetric collisions, several results can be observed: breather
generation, with the formation of a trapped breather and two new
moving breathers; breather reflection; generation of two new moving
breathers; and breather fusion bringing about a trapped breather.
For non-symmetric collisions the possible results are: breather
generation, with the formation of three new moving breathers;
breather fusion, originating a new moving breather; breather
trapping with also breather reflection; generation of two new moving
breathers; and two new moving breathers traveling as a ligand state.
Breather annihilation has never been observed.

\end{abstract}

\end{frontmatter}

\section{Introduction}

 The study of nonlinear localized excitations in lattices of oscillators, which
receive the name of intrinsic localized modes or discrete breathers
\cite{PHYSD06,CHAOS03,PHYSD99,FW98}, is an active research field in
nonlinear physics. These vibrational modes are rather generic in
models of Klein-Gordon and FPU lattices~\cite{MA94,A97,ST88,SJCA04}.
They also appear as solutions of the Discrete Nonlinear
Shr\"{o}dinger (DNLS) equation \cite{PanosDNLS} where these
excitations are usually known as discrete solitons.

Under certain conditions, stationary breathers can be made mobile
\cite{CAT96,AC98}, i.e., when they experience appropriate
perturbations, the breathers travel through the chain and they are
called moving breathers (MBs). There are no exact solutions for MBs,
but they can be obtained by means of numerical calculations. The
conditions for the existence of MBs in Klein-Gordon lattices are
strongly dependent on the exact details of both the on-site and the
interaction potentials. One of the most thoroughly studied
Klein-Gordon models where MBs appear is the Hamiltonian Klein-Gordon
chain with Morse on-site potential and harmonic coupling potential
\cite{CAGR02,CPAR02,CPAR02b}. Variants of this model have been
proposed in the study of the DNA molecule, for example the
Peyrard-Bishop model \cite{PB89,DPW92}.

In a real discrete system, MBs should appear at arbitrary positions,
then, it is natural to be interested in their collisions. The study
of collisions of MBs has been initiated in FPU chains \cite{Doi03}.
However, in Klein-Gordon chains, the studies have been limited to
the interaction of moving low-amplitude breathers with stationary
high-amplitude ones \cite{DP93,FPM94,FCP97}, or to the interaction
between quasi-periodic moving breathers in dissipative
lattices~\cite{MF04}. The study of soliton collisions in
non-integrable DNLS models has not been undertaken until very
recently. These models deal with nearly integrable DNLS
equations~\cite{CBG97,DKMF03}, cubic DNLS equations~\cite{PKMF03}
and saturable DNLS equations~\cite{CE06,MHS06}.

The aim of this paper is to get some insight into the detailed
mechanisms and possible outcomes of collisions in a Klein-Gordon
chain of oscillators with Morse on-site potential. We have
considered only two types of collisions: a)~symmetric collisions,
that is, collisions of two identical MBs traveling with opposite
velocities; b)~non-symmetric collisions, or head-on collisions of
two MBs with the same frequency but different velocities.

This article is organized as follows. Sec.~\ref{sec:model}
introduces the model, describes the means for producing MBs and
different types of collisions. Sec.~\ref{sec:numerical}, presents
the numerical simulations results corresponding to symmetric and
non-symmetric head-on collisions. Sec.~\ref{sec:breather} presents
some plausible explanations for the different outcomes. In
Sec.~\ref{sec:comparison}, we compare the results  for collisions of
discrete solitons in DNLS models with the outcomes of our model. The
summary and conclusions are presented in Sec.~\ref{sec:conclusions}.

\section{Moving breathers and collisions}\label{sec:model}

We consider a one-dimensional Klein-Gordon chain of identical
oscillators. In scaled variables the Hamiltonian is given by:

\begin{equation}\label{ham}
    H=\sum_n\left[\frac{1}{2}\dot u_n^2+V(u_n)+
    \frac{1}{2}\varepsilon(u_n-u_{n+1})^2\right],
\end{equation}

where $u_n$ represents the displacement of the nth oscillator from
the equilibrium position, $\varepsilon$ is the coupling parameter
and $V(u_n)$ is the Morse on-site potential:

\begin{equation}
    V(u_n)=\frac{1}{2}\left(\exp(-u_n)-1\right)^2.
\end{equation}

Time-reversible, stationary breathers can be obtained using methods
based on the anti-continuous limit\cite{MA96}. At $t=0$, $\dot u_n
=0, \forall n$, and the displacements  of a breather centered at
$n_0$ are denoted by $\{ u_{SB,n} \}$. A moving breather $\{ u_{t,n}
\}$ can be obtained with the following initial displacements and
velocities:

\begin{eqnarray}\label{ic}
    u_{MB,n}^0 &=& u_{SB,n}\cos(\alpha (n-n_0)) \nonumber \\
    \dot u_{MB,n}^0 &=& \pm u_{SB,n}\sin(\alpha (n-n_0))\,\,.
\end{eqnarray}

The plus-sign corresponds to a breather moving towards the positive
direction and the minus one, the opposite.  This procedure taken
from the DNLS context \cite{PKMF03,CE06} works as well as the
marginal-mode method \cite{CAT96,AC98} and gives good mobility for a
large range of $\varepsilon$. The parameter $\alpha$ is the
difference of phase between two neighboring oscillators an we will
refer to it as the {\em wave number}. The translational velocity and
the translational kinetic energy of the MB increase with $\alpha$.
We use Eqs.~(\ref{ic}) as initial conditions to integrate the
dynamical equations using a symplectic algorithm \cite{SC94}. The
number of oscillators $N$ is between 100 and 200 with  periodic
boundary conditions.

The study begins generating two MBs with the same frequency, located
initially far apart, traveling in opposite directions. We have
considered two different types of collisions, symmetric and
non-symmetric collisions. For symmetric collisions, both MBs have
initially the same displacements and opposite velocities given by
Eqs.~(\ref{ic}) with the same wave number $\alpha$.

There are two types of symmetric collisions: a) on-site collisions
(OS), if  initially the centers of the breathers are separated by an
odd number of particles, and b) inter-site collisions (IS), if that
number is even \cite{PKMF03,CE06}.

\section{Collision simulations}\label{sec:numerical}

In this section, we present the results of the numerical simulations
for symmetric and non-symmetric collisions. The breather frequency
$\wb$ is below the phonon band because the on-site potential is
soft.  The lowest frequency of the phonon band is equal to the
linear frequency of each isolated oscillator and is given by
$\wo=V''(0)^{1/2}$. For breathers with small amplitude the system is
close to the linear limit, and their frequency is close to $\wo$.
Therefore, $|\wo-\wb|$ is a measure of how far the system is from
the linear regime. Hence, it is natural to think that the collision
scenario should be strongly dependent of the common frequency of the
two MBs. For that reason, we have considered two different breather
frequencies, $\wb=0.8$ and $\wb=0.95$, that represent different
degrees of nonlinearity. We describe below the different scenarios
that appear considering the two types of collisions and the two
frequencies.

\begin{figure}
    \begin{center}
    \includegraphics[width=0.7\textwidth]{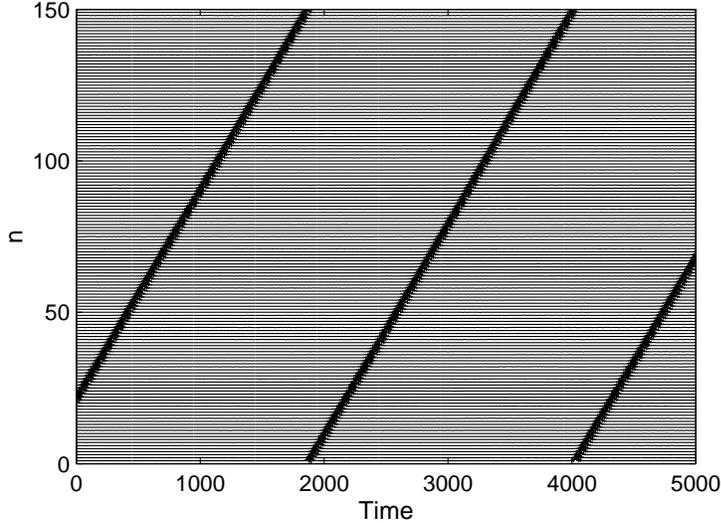}
    \end{center}
    \caption{Displacements versus time in a lattice with 150
    oscillators. Wave number $\alpha=0.1$; coupling parameter $\varepsilon=0.32$ and
breather frequency $\wb=0.8$}.
    \label{Fig1}
\end{figure}

Before considering collisions between MBs, it is necessary to have
an estimate of their lifetime. A MB has not a single frequency but a
continuous band around the frequency of the stationary breather. As
this band contains phonon frequencies, phonons are excited and the
breather loses energy. However, we have checked that in our system
this energy is lost at a small rate and MBs propagate during several
hundreds of periods without apparent decay (see Fig.~\ref{Fig1}). In
our simulations, the time before the collision is something between
20 and 80 periods, therefore, we can consider that the colliding
breathers are almost intact.

\subsection{Symmetric collisions}

We have performed an extensive study of collisions considering
different values of the coupling parameter $\varepsilon$ and MBs
with different values of the wave number $\alpha$. The values of
$\varepsilon$ have been taken in the interval [0.13,0.35] with step
size 0.01. For each value of $\varepsilon$ the values of $\alpha$
have been taken in the interval [0.030,0.200] with  step size 0.002.

Generally, our simulations show that the outcome is strongly
sensitive to the dynamical states of the MBs just before the
collision.

\subsubsection{Symmetric collisions of breathers with frequency $\wb=0.8$}

\begin{figure}
    \begin{center}
    \includegraphics[width=\textwidth]{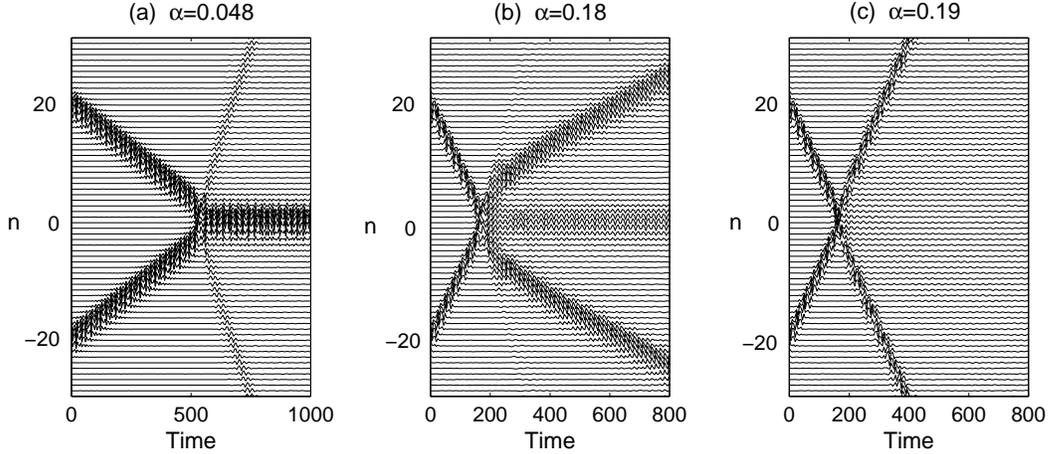}
    \end{center}
    \caption{Three examples of symmetric collisions
    with coupling parameter $\varepsilon=0.32$ and frequency $\wb=0.8$. Displacements
    versus time for three different values of
    the wave number $\alpha$:
    (a) $\alpha=0.048$;  (b) $\alpha=0.18$; (c) $\alpha=0.19$. Note
    that these behaviours occur in an apparently random way when $\alpha$ increases, although
    in these figures they seem to take place progressively.}
    \label{Fig2}
\end{figure}

With this frequency and coupling parameter $\varepsilon=0.32$, there
are no significant differences between the outcomes of OS and IS
collisions. The results can be summarized as follows:

\begin{enumerate}

\item
{\em Breather generation with trapping}:

The collision produces three new breathers, a trapped one located at
the collision region, and two new symmetric MBs, as
Fig.~\ref{Fig2}(a) shows for $\alpha=0.048$. The trapped breather
contains most of the initial energy. This behavior has been
described in the pioneering work cited in Ref.\cite{CAT96}.

 Varying the parameter $\alpha$, it is possible a noticeable
attenuation of the amplitude of the trapped breather, which
anticipates an entirely new outcome. Fig.~\ref{Fig2}(b) corresponds
to $\alpha=0.18$. In this case the emerging MBs contain most of the
initial energy.

\item
{\em Breather reflection}:

\noindent With a slightly different value of $\alpha$, it is
possible a collision which results in two new symmetric MBs, with
almost the same velocity that the colliding breathers'. This is
shown in Fig.~\ref{Fig2}(c) for $\alpha=0.19$.
\end{enumerate}

\begin{figure}
    \begin{center}
    \includegraphics[width=\textwidth]{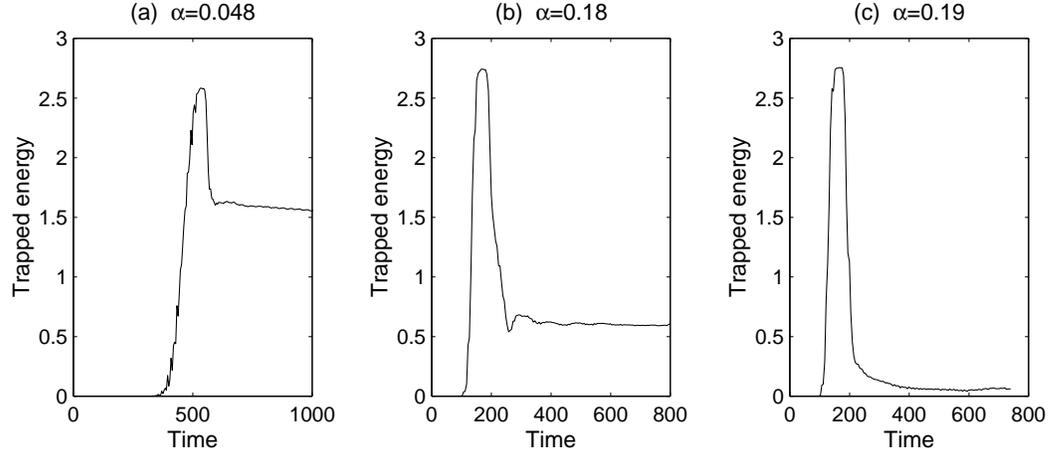}
    \end{center}
    \caption{Time evolution of the central energy corresponding to
    the collisions (a), (b) and (c) of  Fig.~\ref{Fig2},
    respectively.}
    \label{Fig3}
\end{figure}

The total energy transported by the colliding MBs is distributed
after the collision: some part corresponds to the energy of the
trapped breather, another part to the emerging MBs, and a small
fraction of the energy is transferred to the lattice in the form of
phonon radiation. In order to illustrate this phenomenon, we have
studied the evolution of the "central energy", defined in our study
as the energy of eleven particles around the collision region. This
number of particles has been selected because it corresponds to the
typical size of a discrete breather with the parameters used.
Fig.~\ref{Fig3} shows the evolution of the central energy for the
three cases considered in Fig.~\ref{Fig2}.

Before the collision the central energy is zero; after the
initiation of the collision it increases quickly, up to a value very
close to the sum of the incident MBs energies; the subsequent
decrease of the central energy is caused by the appearance of two
emerging MBs and by phonon radiation.

\begin{figure}
    \begin{center}

    \includegraphics[width=\textwidth]{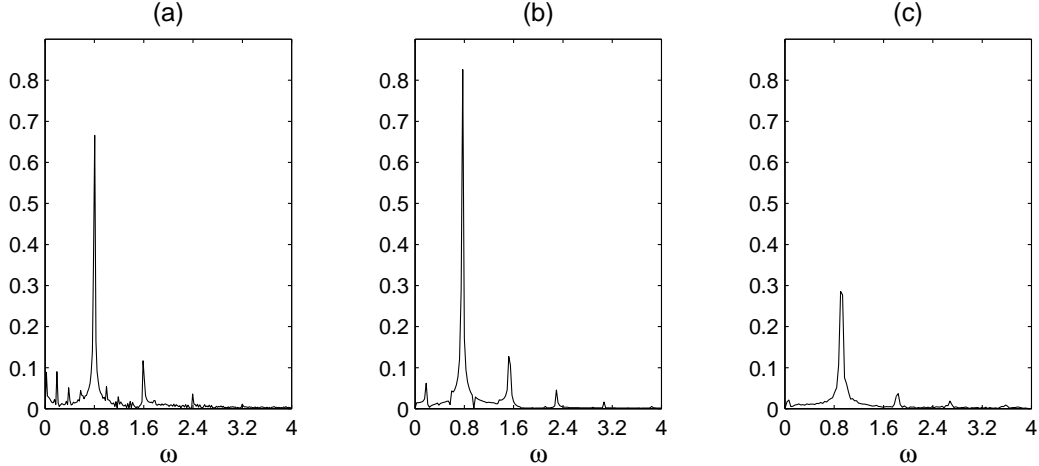}
    \end{center}
    \caption{Fourier spectra of the breathers involved in the
    collisions
    shown in Fig.~\ref{Fig2}(a).
    a) Incident breathers; b) trapped breather; c) emerging breathers.}
    \label{Fig4}
\end{figure}

The Fourier spectra of  the breathers involved in the collisions of
Fig.~\ref{Fig2}(a) are shown in Fig.~\ref{Fig4}. The trapped
breather has a frequency ($\wt\sim 0.77$), which is lower than the
colliding breathers', but its amplitude is larger. The emerging MBs
have higher frequency ($\we\sim 0.90$) and smaller amplitude than
the colliding breathers'. We have also obtained the Fourier spectra
for the breathers shown in Fig.~\ref{Fig2}(b) and (c). For the
former, the trapped and emerging breathers have frequencies higher
than the incident ones' ($\wt=0.90$ and $\we=0.86$, respectively),
but smaller amplitudes. For the latter, the frequency of the
reflected breathers is the same as the incident ones' indicating the
quasi-elastic character of the scattering.

\begin{figure}
    \begin{center}

    \includegraphics[width=\textwidth]{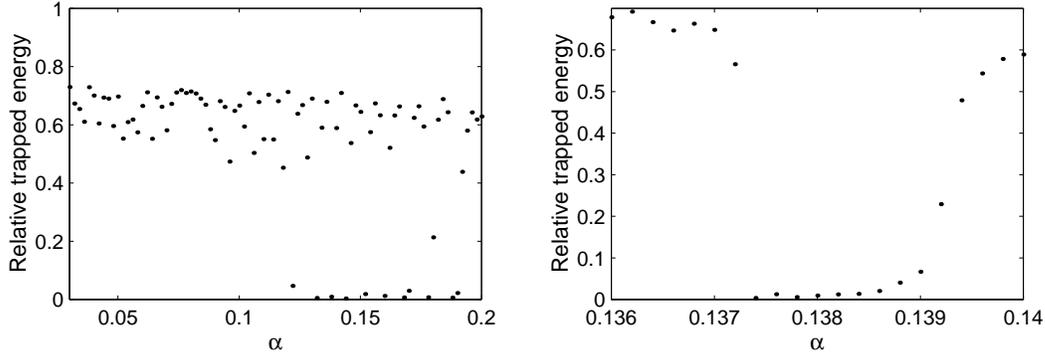}
    \end{center}
    \caption{(Left) Distribution of points representing the relative trapped energy
     versus wave number $\alpha$,  $\alpha\in[0.030,0.200]$ with step size
     0.002. Coupling parameter $\varepsilon=0.32$; breather frequency $\wb=0.8$.
     (Right) Zoom around a value corresponding to breather reflection.}
     \label{Fig5}
\end{figure}

For each set of values $(\varepsilon,\alpha)$, we can calculate the
relative trapped energy in the collision, defined as he ratio
between the energy of the trapped breather and the sum of the
energies of the two incident MBs. Taking $\varepsilon=0.32$ and
$\alpha\in[0.030,0.200]$ with step size 0.002, we have obtained the
relative trapped energies of these collisions which are represented
by the distribution of points shown in Fig.~\ref{Fig5}~(left).

Note that almost all points are distributed inside a band apparently
at random. This means that a small change of $\alpha$ can affects
the energy of the trapped breather. Occasionally, for large enough
values of $\alpha$, there are points outside the band which
correspond to a relative trapped energy close to zero, i.e., for
those values of $\alpha$, the breathers are reflected. The behavior
of the relative trapped energy around one of these points can be
better appreciated if we choose a smaller step size for $\alpha$.
For example, if we take $\alpha$ within the interval [0.1360,0.1400]
with step size 0.0002, we find that for $\alpha=0.1370$ there is
trapping, and for $\alpha=0.1372$ the breathers are reflected.
Fig.~\ref{Fig5}~(right) shows the behavior of the relative trapped
energy around one of these values. There is an abrupt diminution of
the relative trapped energy, that is, breathers reflection appears
as an abrupt process for some exacts values of $\alpha$.

\begin{figure}
    \begin{center}
    \includegraphics[width=0.7\textwidth]{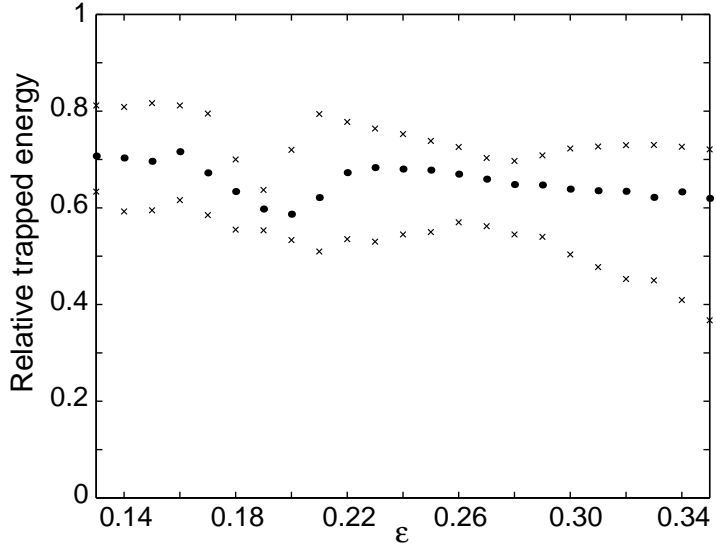}
    \end{center}
    \caption{Dots: dependence of the mean value of the relative
    trapped energies versus $\varepsilon$. Upper x-marks: dependence of the 98th
    percentile of the relative trapped energies versus $\varepsilon$.
    Lower x-marks: dependence of the 2nd percentile of the relative trapped energies
    versus $\varepsilon$.}
  \label{Fig6}
\end{figure}

Varying only the coupling parameter $\varepsilon$, we observe that
the distributions of points are similar although the mean values of
the relative trapped energies and the dispersion of points change.
For each $\varepsilon$ taken in the interval [0.13,0.35] with step
size 0.01, we have obtained the corresponding distribution of points
and we have calculated the mean value, the 98th and the 2nd
percentile of the relative trapped energies. Fig.~\ref{Fig6} shows
the dependence of these quantities versus $\varepsilon$.

\subsubsection{Symmetric collisions of breathers with frequency $\wb=0.95$}

For breather frequencies around this value, which is close to the
frequency of an isolated oscillator in the linear regime, there are
some important differences with respect to the previous case and
there are only two possible outcomes.

\begin{enumerate}
\item {\em Trapping}:
For small enough incoming translational velocities, the outcome is a
bound state of two trapped breathers whose distance oscillates with
a decaying amplitude, i.e., the two trapped breathers have multiple
rebounds, losing energy through phonon radiation and eventually
decaying to a single stationary trapped breather.

\item {\em Reflection}: There exists a critical value, $\alpha_c$, of
the wave number $\alpha$, such that if $\alpha>\alpha_c$ the
breathers are always reflected.
\end{enumerate}

\begin{figure}
    \begin{center}
    \includegraphics[width=0.9\textwidth]{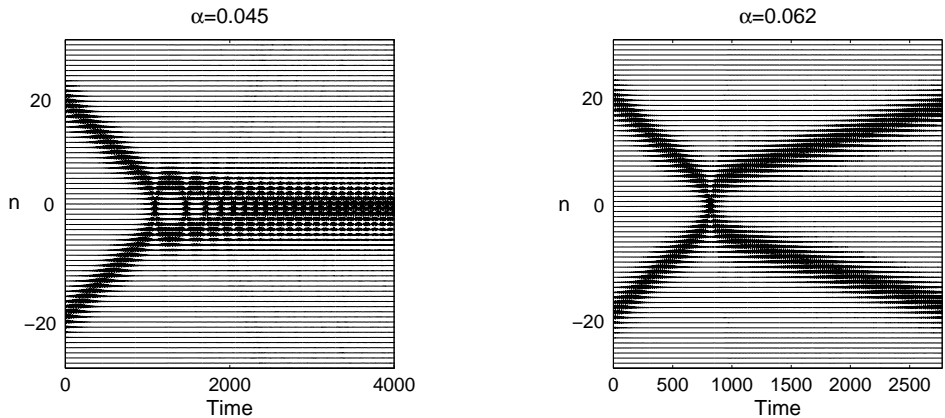}
    \end{center}
    \caption{Two examples of symmetric OS collisions with coupling parameter
    $\varepsilon=0.32$ and breather frequency $\wb=0.95$.
    Displacements versus time for two different values of the wave number $\alpha$.
    (Left) $\alpha=0.045$. (Right) $\alpha=0.062$.}
     \label{Fig7}
\end{figure}

Fig.~\ref{Fig7} shows these two outcomes for OS collisions with
$\varepsilon=0.32$ and $\wb=0.95$. The frequency of the trapped and
reflected breathers of this figure are $\w=0.92$ and $\w=0.95$,
respectively. The scenario for IS collisions is similar.

\begin{figure}
    \begin{center}
    \includegraphics[width=0.7\textwidth]{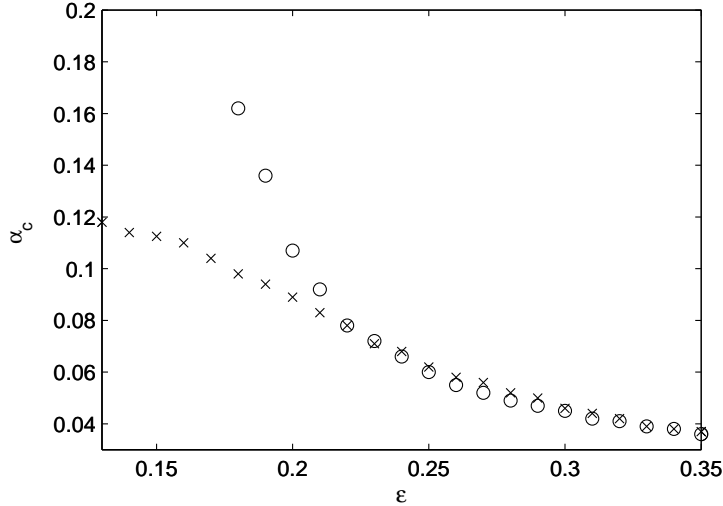}
    \end{center}
    \caption{ Representation of the critical wave number $\alpha_c$ versus
    the coupling parameter $\varepsilon$ for symmetric collisions of breathers with
    frequency $\wb=0.95$.
    Circles correspond to OS collisions and x-marks to IS collisions}
     \label{Fig8}
\end{figure}

The critical value $\alpha_c$ depends on $\varepsilon$, and the
simulations show that for low coupling, the values obtained with OS
collisions are larger  than for IS collisions. Nevertheless, they
approach as $\varepsilon$ increases and they are practically
coincident for $\varepsilon > 0.22$. Fig.~\ref{Fig8} shows the
dependence of $\alpha_c$ versus $\varepsilon$ for IS and OS
collisions.

\subsection{Non-symmetric collisions}

The symmetry is broken if the incoming breathers have different
translational kinetic energies. Clearly, in this case it is
meaningless to distinguish between OS and IS collisions  due to the
different breather velocities. The simulations show that there are
many different outcomes, so we briefly describe some of them.
Hereafter, the two MBs are denoted as MB$_1$ and MB$_2$, and their
wave numbers are represented by $\alpha_1$ and $\alpha_2$,
respectively.

\begin{figure}
    \begin{center}
    \begin{tabular}{cc}
    \includegraphics[width=0.9\textwidth]{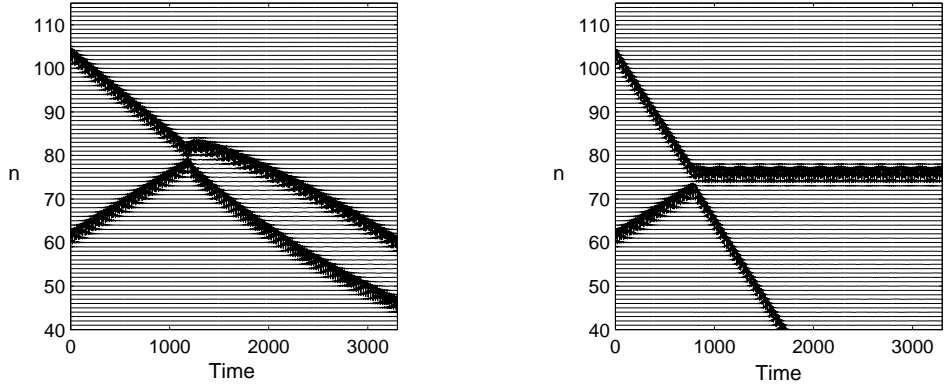}
    \end{tabular}
    \end{center}
    \caption{ Two examples of non-symmetric collisions
    with $\varepsilon=0.15$, $\wb=0.8$, and $\alpha_1=0.042$ with
    different outcomes. (Left)
    Two emerging MBs with the same direction, for $\alpha_2=0.061$.
    (Right) A reflected breather and the generation of a trapped
    breather, with $\alpha_2=0.131$.}
     \label{Fig9}
\end{figure}

\begin{figure}
    \begin{center}
    \begin{tabular}{cc}
    \includegraphics[width=0.9\textwidth]{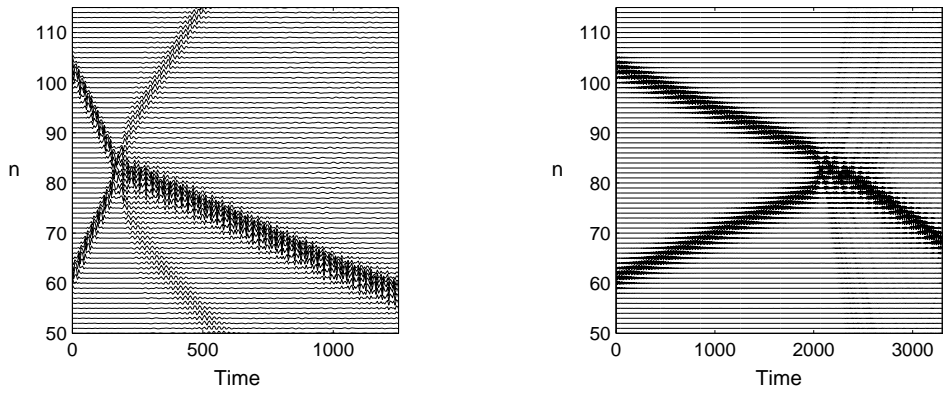}
    \end{tabular}
    \end{center}
    \caption {(Left) A non-symmetric collision with the generation of
    three new MBs of different amplitudes, with $\varepsilon=0.32$, $\wb=0.8$, $\alpha_1=0.18$,
    and $\alpha_2=0.181$. (Right) Merging of two colliding MBs into a single MB, with $\wb=0.95$,
     $\varepsilon=0.14$, $\alpha_1=0.048$, and $\alpha_2=0.046$.}
     \label{Fig10}
\end{figure}

\begin{enumerate}
\item {\em Collisions of breathers with frequency $\wb=0.8$}

\begin{enumerate}
\item {\em Small coupling}:

There are many different outcomes. The simulations selected to
present this case correspond to
 $\epsilon=0.15$, $\alpha_1=0.042$, and $\alpha_2\in[0.031,0.2]$
 with step size $0.001$.
\begin{enumerate}
\item For $\alpha_2<\alpha_1$, i.e.,  MB$_1$ moves faster than MB$_2$, we have observed two main behaviours:
1) only a slow MB emerges traveling in the direction of MB$_2$; 2) a
breather is trapped at the collision region with or without the
appearance of two outgoing MBs.

\item For $\alpha_2>\alpha_1$, i.e.,  MB$_2$ moves faster than MB$_1$, we have observed three main
behaviours: 1) two MBs of different amplitudes emerge with different
velocities and opposite directions, or with the same direction of
MB$_2$ (in this case, the phonons can help to the formation of a
ligand state, as shown in Fig.~\ref{Fig9}--left); 2) a single
trapped breather and a single MB of small amplitude traveling in the
direction of either MB$_1$ or MB$_2$ (see Fig~\ref{Fig9}--right); 3)
a slow MB in the direction of MB$_1$, and two emerging MBs with
smaller amplitudes.
\end{enumerate}
\item {\em Large coupling}:

For large coupling the behaviours  are rather different. Taking
$\varepsilon=0.32$, we have found either two MBs traveling in
opposite directions or a MB with two MBs of small amplitude
traveling in opposite directions (see Fig.~\ref{Fig10}--left). We
have not found any of the other results observed with low coupling.
\end{enumerate}

\item {\em Collisions of breathers with frequency $\wb\sim 0.95 $}

If the breather frequency approaches to the bottom of the phonon
band, i.e., the system is close to the linear regime, the asymmetric
collisions almost always produce two reflected MBs. Nevertheless, at
low coupling and small enough breathers velocities, the two MBs can
merge originating a new MB. This occurs when the two MBs have almost
the same velocities. Fig.~\ref{Fig10}--right illustrates a collision
with $\wb=0.95$, $\epsilon=0.14$, $\alpha_1=0.048$ and
$\alpha_2=0.046$.
\end{enumerate}

\section{Breather stability and trapping}\label{sec:breather}

There are some possible explanations for the trapping mechanism
based in the stability properties of stationary breathers.

Consider a symmetric collision resulting in breather generation, as
shown in Fig.~\ref{Fig2}~(a). The internal and  kinetic energies of
the breathers involved and the energy emitted through phonon
radiation are related by

\begin{equation}\label{eq:enbal}
   2\Ui+ 2\Ki=\Ut+ 2\Uo+ 2\Ko+ \Uph,
\end{equation}
where $\Ui$ and $\Ki$ represent the internal and kinetic energies of
each one of the incident breathers; $\Uo$ and $\Ko$ represent the
internal and kinetic energies of each one of the emerging breathers;
$\Ut$ and $\Uph$  represent, the internal energy of the trapped
breather and the energy emitted through phonon radiation during the
collision, respectively.

\begin{figure}
\begin{center}
    \begin{tabular}{cc}
    \includegraphics[width=0.48\textwidth]{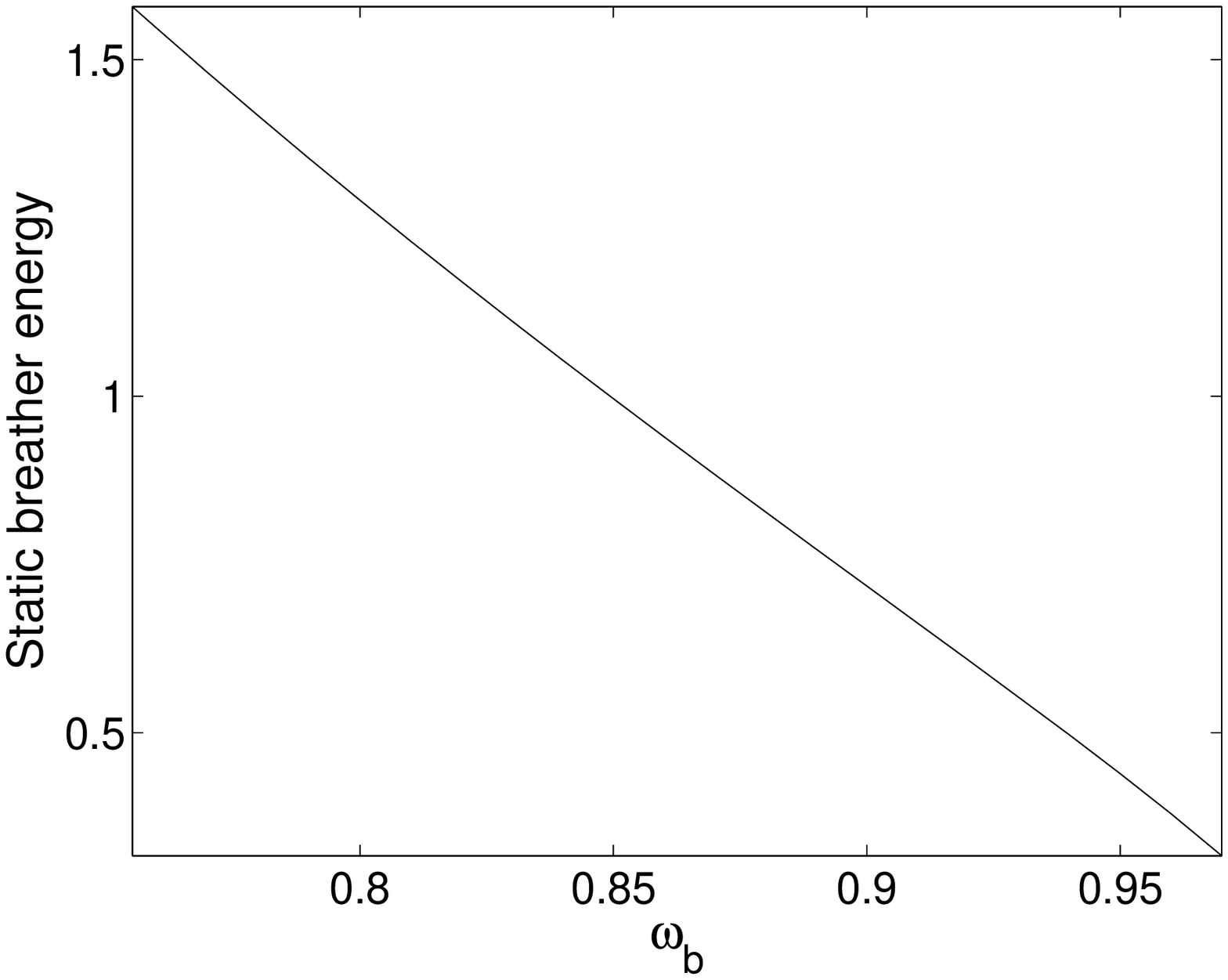}&
    \includegraphics[width=0.48\textwidth]{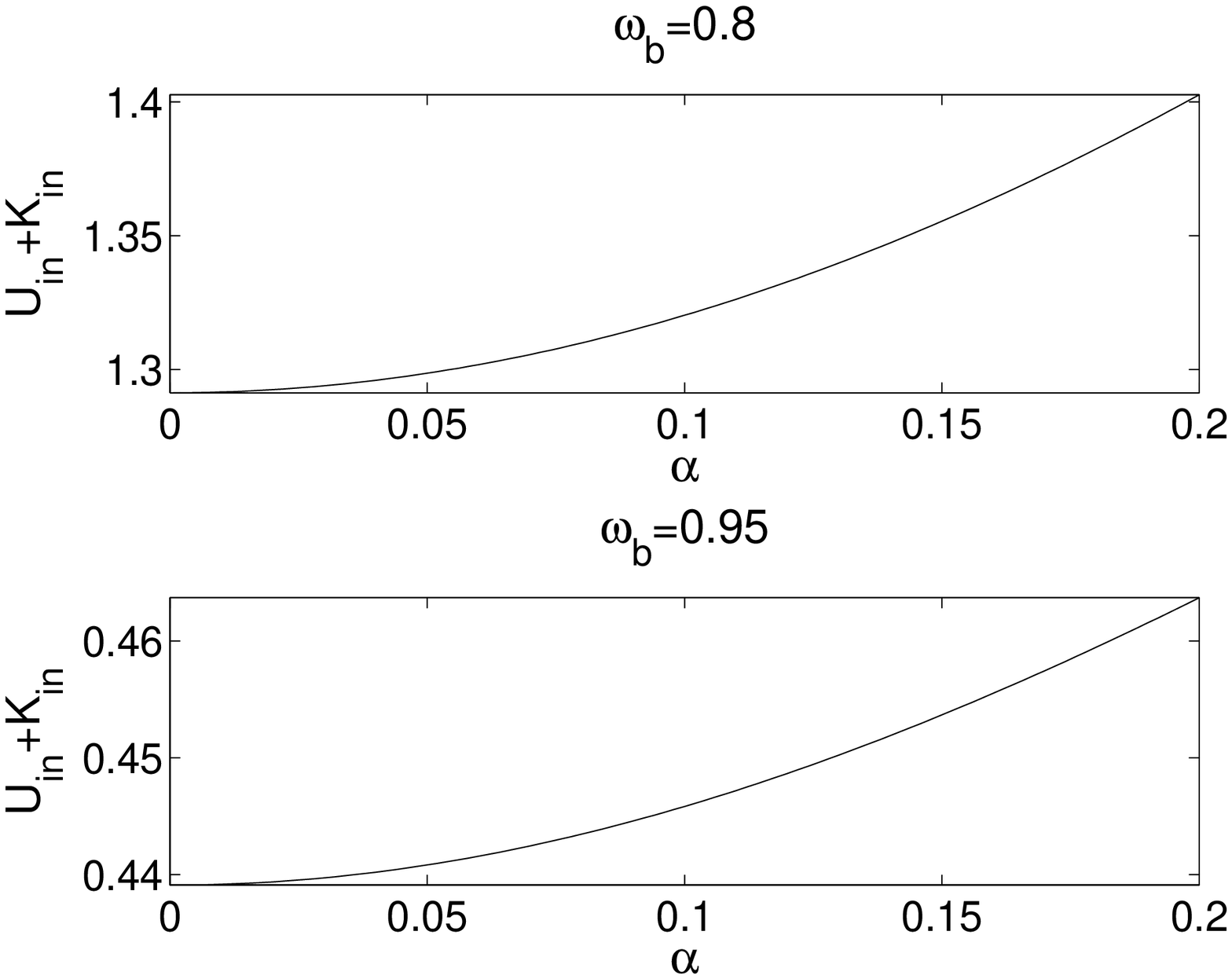}
    \end{tabular}
\caption{ (Left) Dependence of the energy of a stationary breather
with respect to $\wb$ with $\epsilon=0.32$. The maximum value of the
energy is $\tilde E=1.5798$ and corresponds to
$\wb=\sqrt{0.25+\epsilon}=0.7550$ (resonance of $2\wb$ with the
upper edge of the phonon band). (Right) Sum of the internal and
kinetic energies versus $\alpha$ for breathers with $\epsilon=0.32$
and $\wb=0.8$ (top), or $\wb=0.95$ (bottom). }
 \label{fig11}
\end{center}
\end{figure}

The on-site potential of our Klein-Gordon model is soft and, for
this type of potential, an increase of the breather frequency
corresponds to a decrease of the internal energy of the breather.
The energy of the stationary breather versus $\wb$, with
$\epsilon=0.32$, is shown in Fig.~\ref{fig11}--(left). The frequency
of a trapped breather is always different from the frequency of the
incoming breathers, as the Fourier spectra  shows. The generation of
a trapped breather requires an amount of energy $\Ut$, approximately
equal to the energy of a stationary breather with the same
frequency. Then, if the trapped breather has a frequency lower than
the frequency of the incoming breathers, $\Ut>\Ui$, and viceversa.

The sum of the internal and kinetic energies of a MB depends on
$\alpha$, $\epsilon$ and $\wb$. Fig.~\ref{fig11}--(right) shows the
dependence of this sum  with respect to $\alpha$ for MBs with
$\epsilon=0.32$ and $\wb=0.8$ (top), or $\wb=0.95$ (bottom).

Considering the collision shown in Fig.~\ref{Fig2}~(a), the
frequencies of the trapped and emerging breathers are $\wt=0.77$ and
$\we=0.90$, respectively.
The following approximate results hold
   $2\Ui+ 2\Ki= 2.6$, $2\Uo+ 2\Ko= 1$, $\Ut= 1.5$ and $\Uph= 0.1$,
   which means that about $3.8\%$ of the incident energy is
   lost as phonon radiation. However, for the collision shown in
   Fig.~\ref{Fig7}--left about $30\%$ of the
   energy is lost as phonon radiation.

As we have shown, when two MBs collide, the excited region emits
phonon radiation and the oscillators have a small frequency shift.
For symmetric collisions, this shift depends on $\alpha$, $\epsilon$
and $\wb$. If a stationary breather with this new frequency is
stable, a trapped breather appears and the remaining energy is
emitted as two new MBs traveling with opposite directions (see
Fig.~\ref{Fig2}~(a)~,~(b)) or as additional phonon radiation (see
the multiple rebounds in Fig.~\ref{Fig7}--left). However, if for the
new frequency the stationary breather is unstable, there is no
trapping at all and two new MBs appear traveling with opposite
directions (see Fig.~\ref{Fig2}(c) and Fig.~\ref{Fig7}-right).

The stability of a breather can be studied by means of its Floquet
eigenvalues. Considering a breather solution $\{u_n(t)\}$ with
period $T$, if $\xi(t)$ and $\pi(t)$ represent a perturbation of the
positions and velocities with respect to $\{u_n(t)\}$, the Floquet
matrix $\F$ is defined as
\begin{equation}
\left[ \begin{array}{c}\xi(T)\\ \pi(T)\end{array}\right]=\F
\left[ \begin{array}{c}\xi(0)\\ \pi(0)\end{array}\right]\, .
\label{eq:Floquet}
\end{equation}

The Floquet matrix can be obtained by numerically integrating the
perturbation equations:
\begin{equation}
  \ddot{\xi}_n(t)+V''(u_n(t))\cdot\xi_n(t)+
  \varepsilon(2\,\xi_n(t)-\xi_{n+1}(t)-\xi_{n-1}(t))=0\, .
\label{eq:perturbations}
\end{equation}

The $2 N$ eigenvalues of $\F$, $\{\lambda_i\}$, are called the
Floquet multipliers. They can be expressed as $\lambda_i=\exp(\ii
\theta_i)$, where the complex numbers $\{\theta_i\}$ are called the
Floquet arguments.

The perturbation equations are real and symplectic, which implies
\cite{A97} that if $\lambda_i$ is a multiplier, the complex
conjugate $\lambda_i^*$, the inverse $1/\lambda_i$ and
 $1/\lambda_i^*$ are also multipliers. A perturbation parallel to $\dot{u}_n$ is
also solution to the perturbation equations and it is called the
{\em phase mode} because it represents a change in the phase. As it
is periodic, its multiplier has modulus 1 and, therefore, there is
always a double 1 among the Floquet multipliers.

A breather is stable if every multiplier satisfy
 $|\lambda_i |\leq 1$, but if, for some $i$, $|\lambda_i |<1$, then $|1/\lambda_i |>1$.  Therefore the
stability condition is that every eigenvalue has modulus 1 (i.e., it
belongs to the unit circle), or, equivalently, every Floquet
argument is a real number.

If $\{u_n(t)\}$ corresponds to a stable breather with frequency
$\wb$, all the multipliers belong to the unit circle. If the
frequency changes, the multipliers move along the circle, except the
double 1 corresponding to the phase mode. The breather becomes
unstable when the multipliers leave the circle and a stability
bifurcation takes place. There are only three different bifurcation
types:

 a)~Harmonic bifurcation: two multipliers coincide at 1 and
leave the unit circle as real positive numbers, one smaller and the
other larger than 1.

b)~Subharmonic bifurcation: two multipliers coincide at -1 and leave
the unit circle as real negative numbers, one smaller and the other
larger than -1.

c)~Oscillatory bifurcation: two complex eigenvalues collide  and
leave the unit circle as complex numbers (and their complex
conjugates).

A further constraint for the appearance of a bifurcation is that the
Krein signatures of the multipliers that are going to leave the unit
circle must have different signs~\cite{A97}. The Krein signature
$\kappa(\lambda_i)$ of a complex multiplier $\lambda_i$ with
eigenvector $[\{\xi_n^i\},\{\pi_n^i\}]$ is defined as:

\begin{equation}
\kappa(\lambda_i)=\mathrm{sign}\,
\left(\sum_n\ii\,[\xi^i_n(t)\pi^{i*}_n(t)- \xi^{i*}_n(t)\pi^i_n(t)]
\right) \, , \label{eq:ksignature}
\end{equation}

which does not change with time due to the
symplectiness of Eqs.~(\ref{eq:perturbations}). The Krein signature
$\kappa(\lambda_i)$ of a real multiplier is zero.

If $u_n=0,\,\forall n$, the solutions of
Eqs.~(\ref{eq:perturbations}) are the phonons given by $\xi_n^{\pm
q}=\exp[\pm \ii\,(\wph(q)\,t-q\,n)]$ with frequencies
$\wph(q)=[\wo^2+4\,\varepsilon \sin^2(q/2)]^{1/2}$ and wave numbers
$q=2\pi\,m/N$, $m=0,\dots,N-1$. The values of these solutions and
their derivatives at $t=0$, i.e.,
$[\{\xi^\pm_n(0)\},\{\dot{\xi}^\pm_n(0)\}]$ are eigenvectors of the
Floquet matrix, with multipliers $\lambda(\pm q)=\exp(\pm \ii
\wph(q)\,T)= \exp(\pm \ii 2\pi\wph(q)/\wb)$, arguments $\theta(\pm
q)=\pm 2\pi\wph(q)/\wb$ (mod $2\pi$), and Krein signatures
$\kappa(\pm q)=\pm 1$. We will call them, for short, the {\em phonon
multipliers} and {\em phonon arguments}.

If $\{u_n\}$ corresponds to a breather solution, most of the Floquet
multipliers will be almost equal  to the phonon ones, and more so,
the larger the system. They will form two phonon bands within the
unit circle, one with Floquet arguments between $2\pi\wo/\wb$ and
$2\pi[\wo^2+4\,\varepsilon \sin^2(q/2)]^{1/2}/\wb$  with Krein
signature $+1$, and the corresponding complex conjugate band with
Krein signature $-1$. For example, for $\varepsilon=0.19$ and
$\wb=0.8$, the phonon band "+" (with positive Krein signature)
extends from $90^\circ$ to $237^\circ$, the phonon band "-" extends
from $-90^\circ$ to $-237^\circ$. That is, the two phonon bands
overlap and there are arguments with different Krein signature very
close one to each other. Note that we have used in our simulations
around 100 or 200 oscillators, therefore, there are nearly 100 or
200 Floquet arguments in each phonon band.

If $\wb$ changes, the phonons Floquet arguments move along the
circle and many with different Krein signatures will cross, bringing
about the possibility of instabilities. If we calculate numerically
the Floquet arguments as a function of $\wb$, we can observe that
usually they leave the unit circle (i.e. the breather becomes
unstable) and come back again inside it. There are very many
frequency islands of instability and stability. Hence the extreme
sensitivity of the outcome of the collision to the initial
conditions. Moreover, not only the stability of the candidate to
trapped breather changes but also the eigenvector corresponding to
that instability and, as a consequence, the particular result of the
non--trapping collision. At present we can not predict the small
frequency shifts resulting of a symmetric collision, and therefore,
its outcome.

\section{Comparison with discrete soliton collisions in the DNLS equation}\label{sec:comparison}

The scenarios for breather collisions in Klein-Gordon lattices can
be compared with the scenarios for soliton collisions in
non-integrable DNLS models. For these models, the known results deal
with nearly integrable discretizations of the NLS
equation~\cite{CBG97,DKMF03}, cubic DNLS equations~\cite{PKMF03} and
saturable DNLS equations~\cite{CE06,MHS06}.

To start with, we consider the frequency $\wb=0.95$. For frequencies
close to 1, breathers in Klein-Gordon lattices approximate to
envelope discrete solitons of the DNLS equation (see e.g.
\cite{FPM94,BP95,MJKA02}). Then, for the comparison we have selected
the cubic DNLS equation \cite{PKMF03}, where a semi-analytical
variational approximation correctly predicts the main features of
the collisions.

For the cubic DNLS equation there exists a transmission and a
merging regime for both OS and IS collisions (similar to our
reflection and trapping regimes), and there is also a critical value
for the wave number $\alpha$ that separates both regimes.
Nevertheless, in the DNLS case there is a significant difference
between the critical values for OS and IS collisions, which is
explained by the existence of a high Peierls-Nabarro (PN) potential
induced by the lattice. One of the most noteworthy phenomenon that
appear for IS collisions is the possibility of bouncing after
multiple collisions. For both types of collisions, symmetry breaking
is possible, its strongest manifestation being a merger of a pair of
symmetric solitons into a single moving soliton.

\begin{figure}
\begin{center}
    \begin{tabular}{cc}
    \includegraphics[width=0.48\textwidth]{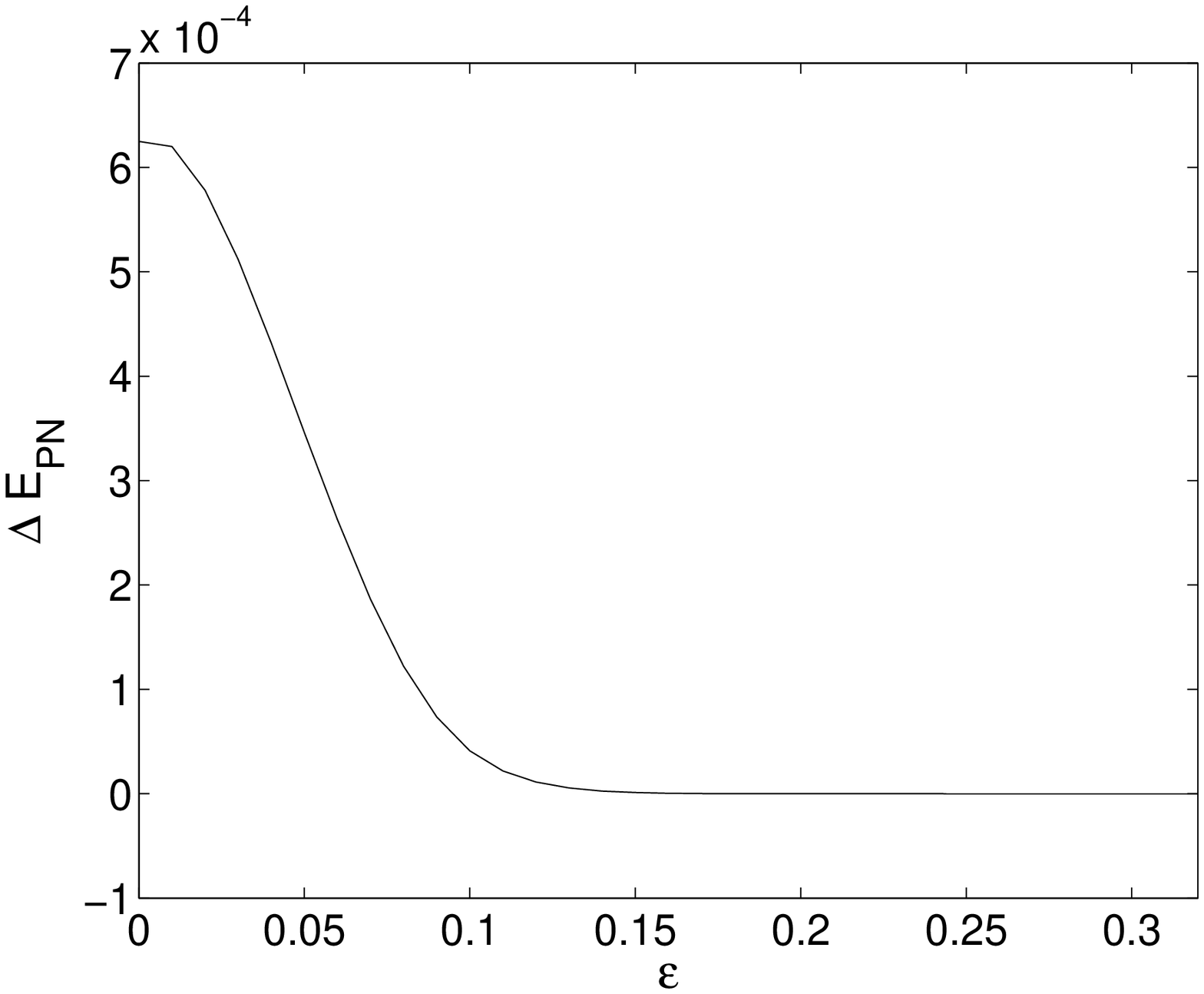}&
    \includegraphics[width=0.48\textwidth]{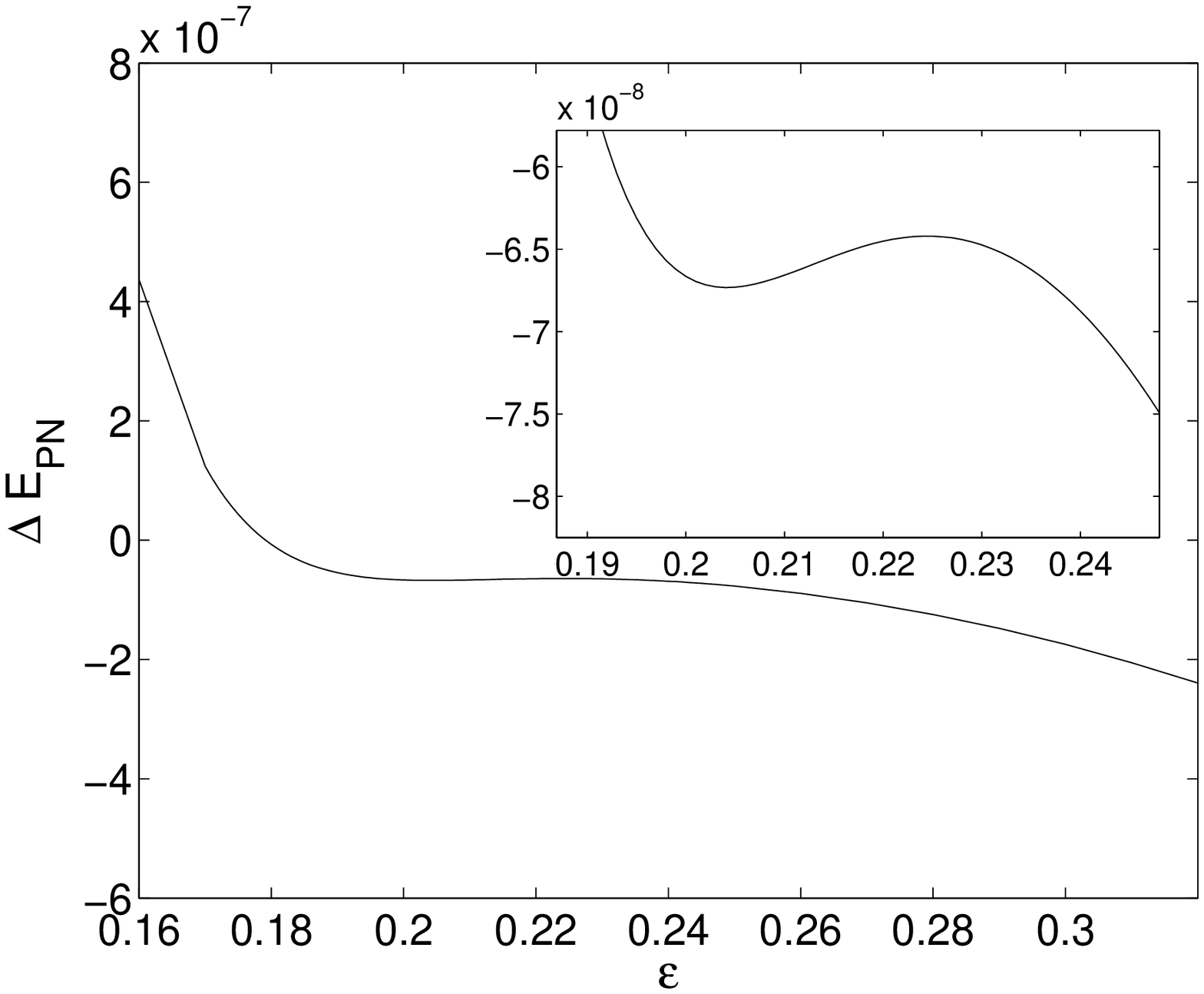}
    \end{tabular}
\caption{ (Left) Representation of the PN barrier versus the
coupling parameter $\varepsilon$ with breather frequency $\wb=0.95$.
(Right) Two zooms of the left figure.}
 \label{Fig12}
\end{center}
\end{figure}

The scenario in the Klein-Gordon chain is quite simple: we have
neither found symmetry--breaking effects, nor bouncing after
multiple collisions, and, contrary to the DNLS case, there is no
qualitative differences between the OS and IS collisions. An
explanation for this similitude can be found studying the PN barrier
in our model. In Klein-Gordon lattices, the PN barrier has been
defined as the difference between the energies of a two-site
breather and a one-site breather, both with the same action
\cite{SEP03}. Fig.~\ref{Fig12}--(left) shows the dependence of the
PN barrier with respect to $\varepsilon$ for $\wb=0.95$, note that
the PN barrier is very small (($\sim10^{-7}$), which is the reason
of the similitude between OS and IS collisions. There is a local
minimum of the PN barrier in the interval
$\varepsilon\in(0.20,0.21)$, as Fig.~\ref{Fig12}--(right) shows.

Finally, we consider the frequency $\wb=0.8$. For the DNLS
equations, there is no formation of ligand states, there are only
breather generation and breather reflection, and there is not a
critical value of the wave number $\alpha$ separating these regimes.
However, the DNLS equation with saturable nonlinearity
\cite{CE06,MHS06} provides some similitudes, as there are reflection
and generation regimes separated by a critical value of $\alpha$.
But in this DNLS case breather generation occurs only when $\alpha$
is larger than the critical value. The generation of ligand states
is also possible.

\section{Conclusions}\label{sec:conclusions}

In this work, we have analyzed collisions of MBs in a Klein-Gordon
model of oscillators. We have studied both symmetric and
nonsymmetric head-on collisions between breathers of the same
frequency. We have considered different values of the coupling
parameter, different velocities and different frequencies of the
colliding MBs.

For symmetric collisions, we have found several scenarios: breather
generation, with the formation of a trapped breather and two new
moving breathers; breather reflection; generation of two new moving
breathers; and breather fusion originating a trapped breather. For
non-symmetric collisions the observed results are: breather
generation, with the formation of three new moving breathers;
breather fusion, originating a new moving breather; breather
trapping with breather reflection; generation of two new moving
breathers; and two new moving breathers traveling as a ligand state.
We have never observed breather annihilation.

For low enough frequency the outcome is strongly dependent of the
velocities of the incident breathers and of their dynamical states
when the collision begins. Very small changes of the velocities can
determine an entire new outcome. This sensitivity disappears for
frequencies close to the frequency of an isolated oscillator in the
linear regime.

 Some additional simulations are underway, these are: head-on
collisions of two breathers with different frequencies and equal or
different velocities; collisions of MBs with a stationary breather
with equal or different frequencies; and collisions of two MBs
traveling in the same direction with equal or different frequencies.

\section*{Acknowledgements}

We acknowledge financial support from the MECD/FEDER project
FIS2004-01183. We are also indebted to Panayotis G. Kevrekidis for
his useful comments.

\newcommand{\noopsort}[1]{} \newcommand{\printfirst}[2]{#1}
  \newcommand{\singleletter}[1]{#1} \newcommand{\switchargs}[2]{#2#1}

\end{document}